\documentclass[preprintnumbers,amsmath,amssymb,usenatbib,nofootinbib,superscriptaddress,showkeys,showpacs,altaffilletter,11pt]{revtex4-1}
\usepackage{graphicx}% Include figure files
\usepackage{dcolumn}% Align table columns on decimal point
\usepackage{bm}% bold math
\usepackage{latexsym}
\usepackage{subfigure}
\usepackage{mathrsfs}
\usepackage{tabu}
\usepackage{color,soul}
\usepackage{amsmath}
\usepackage{amsfonts,color}
\usepackage{amssymb,float}
\usepackage{graphicx}
\usepackage{appendix}
\usepackage[colorlinks]{hyperref}
\usepackage{xcolor}
\usepackage{orcidlink}
\usepackage{epsf}
\usepackage{bm}
\usepackage{epstopdf}
\usepackage{natbib}
\usepackage{lipsum}

\begin{document}

\title{Observational constraints and stability in viscous $f(T,\mathcal{T})$ gravity}

\author{T. Mirzaei Rezaei}
\affiliation{Department of Physics, Ayatollah Amoli Branch, Islamic Azad University, Amol, Iran}

\author{Alireza Amani\orcidlink{0000-0002-1296-614X}}
\email{a.r.amani@iauamol.ac.ir}
\affiliation{Department of Physics, Ayatollah Amoli Branch, Islamic Azad University, Amol, Iran}

\author{E. Yusofi}
\affiliation{Department of Physics, Ayatollah Amoli Branch, Islamic Azad University, Amol, Iran}

\author{S. Rouhani}
\affiliation{Department of Physics, Science and Research Branch, Islamic Azad University, Tehran, Iran}

\author{M. A. Ramzanpour}
\affiliation{Department of Physics, Ayatollah Amoli Branch, Islamic Azad University, Amol, Iran}

\date{\today}

\begin{abstract}
In this paper, we study the $f(T,\mathcal{T})$ gravity model in the presence of the bulk viscosity by the flat-FRW metric. The field equation is obtained by teleparallel gravity with tetrad field. The universe components are considered as matter and dark energy which the dark energy component associates from the viscous $f(T,\mathcal{T})$ gravity. After calculating the Friedmann equations, we obtain the energy density, the pressure and the EoS of dark energy in terms of the redshift parameter. Afterward, we plot the corresponding cosmological parameters versus the redshift parameter and examine the accelerated expansion of the universe. In the end, we explore the system stability by a function called the speed sound parameter.
\end{abstract}

\pacs{98.80.-k; 98.80.Es; 95.36.+x}

\keywords{Dark energy; Equation of state parameter; The $f(T, \mathcal{T})$ gravity;  Bulk viscosity, Observational constraints.}

\maketitle

\section{Introduction}\label{s1}
As we know, the discovery of the mysterious energy called dark energy was first studied in type Ia supernovae (SNe Ia) and afterward other studies were done with further data \cite{Riess_1998, Perlmutter_1999}. It has also been studied from other points of view, such as cosmic microwave background and large scale structure \cite{Bennett_2003, Tegmark_2004}. All the aforesaid studies were based on that the dark energy is responsible for describing the accelerated expansion of the universe and requires strong negative pressure. Indeed, about three-quarters of the total energy of the universe is dark energy. One of the particular cosmological parameters that can describe the accelerated expansion of the universe is called the equation of state (EoS) and is equal to proportion the pressure to the energy density. It is worth mentioning that the expansion of the universe is while accelerating, in which the EoS parameter is less than $-\frac{1}{3}$ and in accordance with observational data, the value of the EoS parameter is close to $-1$. For this purpose, various papers have been cited to describe dark energy on a numerous topics such as cosmological constant \cite{Weinberg-1989, Caldwell-2002, Sadeghi1-2009, Setare-2009}, quintessence\cite{Battye-2016, Li-2012, Khurshudyan-2014, Khurshudyan-2015}, phantom \cite{Sadeghi-2013, Sadeghi-2014, Pourhassan-2014, MKhurshudyan11-2015,JSadeghi-2015, MKhurshudyan1-2015}, quintom \cite{Guo-2005, Setare1-2009, Amani-2011}, tachyon \cite{Iorio-2016, Faraoni-2016}, Chaplygin gas models \cite{Amani-2013, Amani-2014, Naji-2014, KhurshudyanJ2-2014, Amaniali-2013, PourhassanKa-2014, PourhassanB-2016}, modified gravity \cite{Khurshudyan1-2014, Sadeghi1-2016, Sadeghi2-2016, Wei-2009, Amani1-2011, Li-2004, Campo-2011, Hu-2015}, holography \cite{Fayaz-2015, Saadat1-2013, KhurshudyanJ-2014, KhurshudyanB-2014, Morais-2017, Zhang1-2017, Bouhmadi1-2016}, new agegraphics \cite{KhurshudyanJ1-2015, SadeghiKhurshudyanJ-2014, SadeghiKhurshudyanJ1-2014, Sadeghi-2009}, bouncing theory \cite{Sadeghi-2010, Amani-2016, Singh-2016}, teleparallel gravity \cite{Sahni-2003, pourbagher1-2020} and braneworld models \cite{Setare-2008, Brito-2015}.

Among the aforesaid models, the teleparallel gravity is an interesting issue that one can describe dark energy with a different approach beyond general relativity. This means that we use a torsion term as torsion scalar $T$ in Einstein-Hilbert action for teleparallel approach rather than the curvature term in general relativity and one so-called teleparallel equivalent of general relativity. Teleparallel gravity was first introduced by Einstein for the unification of electromagnetism and gravity as well as one provided a particular mathematical structure called distant parallelism \cite{Einstein_1928}. This mathematical structure is presented with the tetrad field and used by Weitzenb\"{o}ck connection rather than the Levi-Civita connection in the general relativity. Then, Weitzenb\"{o}ck connection explains the torsion of a manifold geometry without curvature, but in contrast, the Levi-Civita connection can describe the curvature of a manifold geometry without torsion \cite{Weitzenbock_1923, Bengochea_2009, Tamanini_2012}. If the corresponding action, $T$ changes to the function $f(T)$, we are entering a new model called the modified teleparallel gravity in which $f(T)$ is introduced as a function of $T$ and one obtains from the Friedmann equations after applying the observational constraints \cite{Linder_2010, Myrzakulov_2011}.

In what follows, inspired by $f(R, \mathcal{T})$ gravity model in which $\mathcal{T}$ is the trace of the matter energy-momentum tensor, we insert torsion scalar $T$ instead to curvature $R$ to obtain $f(T, \mathcal{T})$ gravity model \cite{Harko_2011, Harko1_2014, Rezaei_2017, Pace_2017, Ghosh_2020}. We remember in standard general relativity that the left side of Einstein's equation only contains space-time geometry, and the right side corresponds to the effect of matter. But in the $f(T, \mathcal{T})$ gravity model, we will see that there is a coupling between matter and geometry in both sides of the field equation, which gives rise to another approach as an alternative for a source of dark energy in modern cosmology. We note that other literatures presented as $f(R,T)$ gravity in which curvature is coupled with torsion scalar \cite{Myrzakulov_2012, Amani1-2015, Amani-2015}.

As we know, for modeling of the dominated matter in the standard cosmology is using the perfect fluid in which has no viscosity. This means that one is in equilibrium and does not produce friction due to heat, and has a reversible dynamic. Therefore, whenever we deviate from the local thermodynamic equilibrium, the perfect fluid does not correspond to the corresponding system and the bulk viscosity fluid replaces it. In this case, in the effect of fluid expansion within the universe, the corresponding system does not have enough opportunity to return to the equilibrium point, this in itself causes that an effective pressure called the bulk viscosity is involved in restoring the system to its thermal equilibrium. Therefore, the dominated bulk viscosity arises the concept of the accelerating expansion of the universe due to the lack of thermal equilibrium. In other words, the bulk viscosity is considered as internal friction that one gives rise to convert the particles kinetic energy to heat. This in turn is introduced as one of the possible ideas that can describe the accelerated universe in late time. Nevertheless, in this job, we consider that the universe has as more realistic features, namely the universe dominates with an viscosity fluid. The existence of bulk viscosity causes to change the effective pressure within the universe, which has a negative value. Therefore, bulk viscosity affects the acceleration of the universe and is a reasonable description of dark energy \cite{Zimdahl-1996, Sadeghi_2013, Amaniali-2013}.

Accordingly, the aforesaid issues are the motivation for studying dark energy as viscous $f(T, \mathcal{T})$ gravity. Thus, in this paper, we are going to examine the acceleration of the universe by the modified teleparallel model in the existence of bulk viscosity in the flat-Friedmann-Robertson-Walker (FRW) background. Next, we use the values of Hubble parameter $H$ versus the redshift parameter $z$ that its results come from Refs. \cite{Farooq_2017, Simon_2005, Stern_2010, Moresco_2012, Blake_2012, Font_2014, Delubac_2015, Moresco_2015}. Then, we will use the achievement of Ref. \cite{Pourbagher_2019} as the parametrization function $E(z) = \frac{H^2}{H^2_0}$ in which $H_0$ is the current Hubble parameter.

The current paper is organized as follows:

In Sec. \ref{s2}, the $f(T,\mathcal{T})$ gravity is described from the viewpoint of the tetrad field in the flat-FRW metric. In Sec. \ref{s3}, we write  the energy-momentum tensor in the presence of the bulk viscosity, and then obtain the Friedmann equations. In Sec. \ref{s4}, we reconstruct the viscous $f(T,\mathcal{T})$ gravity in terms of redshift parameter, and correspond with observational data, and then investigate the stability of the model by the sound speed parameter. Finally, in Sec. \ref{s5} we will conclude our results on the viscous $f(T,\mathcal{T})$ gravity.

%$$$$$$$$$$$$$$$$$$$$$$$$$$$$$$$$$$$$$$$$$$$$$$$$$$$$$$$$$$$$$$$$$$$$$$$$$$$$$$$$$$$$$$$$$$$$$$$$$$$$$$$$$$$$$$
%$$$$$$$$$$$$$$$$$$$$$$$$$$$$$$$$$$$$$$$$$$$$$$$$$$$$$$$$$$$$$$$$$$$$$$$$$$$$$$$$$$$$$$$$$$$$$$$$$$$$$$$$$$$$$$
%$$$$$$$$$$$$$$$$$$$$$$$$$$$$$$$$$$$$$$$$$$$$$$$$$$$$$$$$$$$$$$$$$$$$$$$$$$$$$$$$$$$$$$$$$$$$$$$$$$$$$$$$$$$$$$

\section{The foundation of $f(T,\mathcal{T})$ gravity}\label{s2}
In this section, we are going to address the issue of dark energy using the modified teleparallel theory entitled $f(T, \mathcal{T})$ model in which $T$, $\mathcal{T}=\delta^\nu_\mu \mathcal{T}_\nu\,^\mu$ and $\mathcal{T}_\nu\,^\mu$ are torsion scalar, trace of $\mathcal{T}_\nu\,^\mu$ and the matter energy-momentum tensor, respectively. For this purpose, the corresponding action is written in the following form
\begin{equation}\label{action1}
  S=\int e \left(\frac{f(T,\mathcal{T}) }{2 \kappa^2}+ \mathcal{L}_m\right) d^4x,
\end{equation}
where $\kappa^2=8 \pi G$, and $e=det\left( e^i_{\,\,\mu}\right)$ is determinant of vierbein field $e^i_{\,\,\mu}$, and $\mathcal{L}_m$ is the matter Lagrangian density that is only related to vierbein field. Therefore, we can write an orthonormal basis of tetrad (vierbein) $e_i(x^\mu)$ in terms of the tangent space at point $x^\mu$ of a manifold as $e_i . e_j = \eta_{ij}$ which $\eta_{ij}=diag(+1,-1,-1,-1)$. Also, the tetrad field $e^i_{\,\,\mu}$ be related with metric tensor by $g_{\mu \nu}=\eta_{\mu \nu} e^i_{\,\,\mu} e^i_{\,\,\nu}$ which in that case, we have for the basis of tetrad field the relationships as $e_i\,^\mu e^i\,_\nu=\delta_\nu^\mu$ and $e_i\,^\mu e^j\,_\mu=\delta_i^j$. We note that the Greek letters denote space-time components, and the Latin letters present tangent space components.

In this study, we use the Weitzenb\"{o}ck connection rather than  the Levi-Civita connection in the form
\begin{equation}\label{levi1}
 \Gamma _{\mu \nu }^{\lambda }=e_i\,^\lambda \partial_\mu e^i\,_\nu=-e^i\,_\mu \partial_\nu e_i\,^\lambda.
\end{equation}

Now, we write quantities of Ricci Tensor $R_{\mu \nu}$, asymmetry tensor ${{S}_{\rho }}^{\mu \nu }$, torsion tensor ${{T}^{\lambda }}_{\mu \nu }$ and torsion scalar $T$ in new geometry as
\begin{eqnarray}
 &R_{\mu \nu } = {{\partial }_{\lambda }}\Gamma _{\mu \nu }^{\lambda }-{{\partial }_{\mu }}\Gamma _{\lambda \nu }^{\lambda }+\Gamma _{\mu \nu }^{\lambda }\Gamma _{\rho \lambda }^{\rho }-\Gamma _{\nu \rho }^{\lambda }\Gamma _{\mu \lambda }^{\rho }, \label{RST-1}\\
 &{{S}_{\rho }}^{\mu \nu } =\frac{1}{2}\left( {{K}^{\mu \nu }}_{\rho }+\delta _{\rho }^{\mu }{{T}^{\alpha \nu }}_{\alpha }-\delta _{\rho }^{\nu }{{T}^{\alpha \mu }}_{\alpha } \right),\label{RST-2} \\
 &{{T}^{\lambda }}_{\mu \nu }={{\Gamma }^{\lambda }}_{\nu \mu }-{{\Gamma }^{\lambda }}_{\mu \nu }=e_{A}^{\rho }\left( {{\partial }_{\mu }}e_{\nu }^{A}-{{\partial }_{\nu }}e_{\mu }^{A} \right),\label{RST-3}\\
 &T=T^\lambda\,_{\mu \nu} S_\lambda\,^{\mu \nu},\label{RST-4}
\end{eqnarray}
where
\begin{equation}\label{kmunu}
  {{K}^{\mu \nu }}_{\rho }=-\frac{1}{2}\left( {{T}^{\mu \nu }}_{\rho }+{{T}^{\nu \mu }}_{\rho }-{{T}_{\rho }}^{\mu \nu } \right).
\end{equation}

We obtain the following field equation by varying the action \eqref{action1} with respect to the tetrad field as
\begin{eqnarray}\label{fri2}
\begin{aligned}
e_{a}\,^{\alpha}S_{\alpha}\,^{\rho\mu}\left(\partial_{TT}f\, \partial_{\mu}T+\partial_{T \mathcal{T}}f\,\partial_{\mu}\mathcal{T}\right)+ \left[e^{-1}\partial_{\mu}\left(ee_{a}\,^{\alpha}S_{\alpha}\,^{\rho\mu}\right)-e_{a}\,^{\alpha}T^{\mu}\,_{\nu\alpha}S_{\mu}\,^{\nu\rho}\right]\partial_{T}f\\
 +\frac{1}{4}{e_{a}\,^{\rho}f}
 -\frac{1}{2}\left({e_{a}\,^{\alpha}\mathcal{T}_{\alpha}\,^{\rho}+pe_{a}\,^{\rho}}\right)\partial_{\mathcal{T}}f
 =\frac{\kappa^2}{2}e_{a}\,^{\alpha}\mathcal{T}_{\alpha}\,^{\rho}.
\end{aligned}
\end{eqnarray}

In this job, the universe is considered by flat-FRW metric in the following
\begin{equation}\label{ds2}
d{{s}^{2}}=d{{t}^{2}}-{{a}^{2}}(t)(d{{x}^{2}}+d{{y}^{2}}+d{{z}^{2}}),
\end{equation}
where $a(t)$ is scale factor. The tetrad field obtains as \({{e}^{i}}_{\mu }={{e}^{\mu }}_{i}=diag(1,a,a,a)\), and then torsion scalar is obtained as the follows:
 \begin{equation}\label{torsion1}
 T= -6{{H}^{2}}
\end{equation}
where $H=\frac{\dot{a}}{a}$ is the Hubble parameter. In next section, we explore the viscous fluid model for $f(T, \mathcal{T})$ gravity.

%$$$$$$$$$$$$$$$$$$$$$$$$$$$$$$$$$$$$$$$$$$$$$$$$$$$$$$$$$$$$$$$$$$$$$$$$$$$$$$$$$$$$$$$$$$$$$$$$$$$$$$$$$$$$$$$$$$$$$$$$$$$
%##################################################################################################################
%&&&&&&&&&&&&&&&&&&&&&&&&&&&&&&&&&&&&&&&&&&&&&&&&&&&&&&&&&&&&&&&&&&&&&&&&&

\section{Viscous $f(T, \mathcal{T})$ gravity}\label{s3}
In this section, we describe the realistic evolution of the universe  in the presence of a viscous fluid called bulk viscosity. This means that the viscosity of fluid indicates its resistance to flow, so bulk viscosity plays a role in cosmic pressure, which in turn affects the acceleration of the universe. For this purpose, the energy-momentum tensor in the presence of bulk viscosity yields
\begin{equation}\label{Tij1}
\mathcal{T}_i^j=(\rho_{eff} + p_{eff} + p_{b}) u_i u^j - \left(p_{eff} + p_{b}\right)\,  \delta_i^j,
\end{equation}
where $\rho_{eff}$ and $p_{eff}$ are the effective energy density and the effective pressure of fluid within the universe, and $p_{b} = -3 \xi H$ represents the pressure of bulk viscosity which $\xi$ is a positive constant, and the $4$-velocity $u_\mu$ is $u^i$ = (+1,0,0,0) which one is written $ u_i u^j$ = 1. Therefore, the energy-momentum tensor yields
\begin{equation}\label{tau1}
 \mathcal{T}_i^j =diag( \rho_{eff}, -\bar{p}_{eff}, -\bar{p}_{eff}, -\bar{p}_{eff}),
\end{equation}
where $\bar{p}_{eff} = p_{eff} - 3 \xi H $, and trace of energy-momentum tensor is
\begin{equation}\label{mathT1}
\mathcal{T} =\rho_{eff} - 3 \bar{p}_{eff}.
\end{equation}

By inserting Eqs. \eqref{ds2}-\eqref{mathT1} into field equation \eqref{fri2}, we can find the Friedmann equations with the viscous fluid in the following form
\begin{subequations}\label{fried1}
\begin{eqnarray}
 & \frac{1}{2}f - T \partial_T f = \kappa^2 \rho_{eff} + (\rho_{eff} + \bar{p}_{eff}) \partial_\mathcal{T} f,\label{fried1-1}\\
 &-2 \dot{H} \left(\partial_T f + 2 T \partial_{T T} f \right) = (\kappa^2 + \partial_\mathcal{T} f) (\rho_{eff} + \bar{p}_{eff}) + 2 H \dot{ \mathcal{T}} \partial_{T  \mathcal{T}} f,\label{fried1-2}
\end{eqnarray}
\end{subequations}
where the point demonstrates the derivative with respect to time evolution. The interesting point is that if $f(T, \mathcal{T}) = T$ the aforesaid Friedmann equations are converted to its standard form in general relativity.

Now we consider the effective energy density and the effective pressure of a fluid into the universe in the following form
\begin{subequations}\label{rhopeff1}
\begin{eqnarray}
 & \rho_{eff} = \rho_{DE} + \rho_m,\label{rhopeff1-1}\\
 & p_{eff} = p_{DE} + p_m,\label{rhopeff1-2}
\end{eqnarray}
\end{subequations}
where index $DE$ is related to the contribution of the dark energy, and index $m$ is contribution within the universe. This means that the universe components consist of the matter and the dark energy. Here, the mass influence into the universe is considered as non-interacting with the dark energy. In that case, the continuity equation for mass becomes
\begin{equation}\label{dotrho1}
\dot{\rho}_m + 3 H (\rho_m + p_m) = 0,
\end{equation}
where its solution yields
\begin{equation}\label{dotrho2}
\rho_m = \rho_{m_0} a^{-3(1+\omega_m)},
\end{equation}
where $\omega_m = \frac{p_m}{\rho_m}$ and $\rho_{m_0}$ are the equation of state (EoS) of mass and the integral constant, respectively. In order to solve the modified Friedmann equations \eqref{fried1}, we consider a particular form for function $f(T, \mathcal{T})$ so that one consists only by an additive term as $f(T, \mathcal{T}) = g(T) + h(\mathcal{T})$. In order that find an analytical solution, we choose a simple particular form as $ g(T) =  \alpha T$ and $h(\mathcal{T}) =  \beta \mathcal{T}$. The motivation of this choice is that when the coefficient of $\mathcal{T}$ is zero, we will have the teleparallel gravity equivalent to GR, and when the coefficient of $T$ is zero, we will have a modified gravity \cite{Salako-2015}. Nonetheless, by inserting this function $f(T, \mathcal{T})$ into the Eq. \eqref{fried1} we find
\begin{subequations}\label{friedii}
\begin{eqnarray}
 & -\alpha\,T + \beta\,\mathcal{T}= 2 \left( {\kappa}^{2}+\beta \right) \rho_{eff} + 2 \beta \left( p_{eff} - 3\,\xi\,H \right),\label{friedii-1}\\
 & -2 \alpha \dot{H} = \left( {\kappa}^{2}+\beta \right) \left( \rho_{eff} + p_{eff} - 3\,\xi\,H \right).\label{friedii-2}
\end{eqnarray}
\end{subequations}

To substitute Eqs. \eqref{torsion1}, \eqref{mathT1}, and \eqref{rhopeff1} into the aforesaid relationships, we can obtain the energy density and the pressure of dark energy in the following form
\begin{subequations}\label{rhopdei}
\begin{eqnarray}
 & \rho_{DE} = \frac{3 \alpha}{\kappa^2 - 2 \beta} H^2 + \frac{5 \alpha \beta}{\left( {\kappa}^{2}+\beta \right)\left( {\kappa}^{2}-2 \beta \right)} \dot{H} - \rho_m,\label{rhopdei-1}\\
 &p_{DE} =  -\frac {3 \alpha}{\kappa^{2}-2\,\beta} H^2 - \frac {\alpha\, \left( 2 \kappa^{2}+\beta \right)}{ \left(\kappa^{2}+
\beta \right)  \left(\kappa^{2}-2\,\beta \right) } \dot{H} + 3\,\xi\,H - \omega_m \rho_m,\label{rhopdei-2}
\end{eqnarray}
\end{subequations}
where these equations show us that the energy density and the pressure of dark energy associate to $f(T, \mathcal{T})$ gravity and bulk viscosity. It should be mentioned that the free parameters play an important role in the confirmation of the accelerated expansion of the universe. For this purpose, we use the observational constraints and also we must have that the EoS is smaller than $-\frac{1}{3}$. Therefore, the EoS of dark energy is given by
\begin{equation}\label{eos1}
 \omega_{DE} = \frac{p_{DE}}{\rho_{DE}}.
\end{equation}

In the next section, we will continue the aforesaid approach to study on a reconstruction by redshift parameter and observational constraints.

%$$$$$$$$$$$$$$$$$$$$$$$$$$$$$$$$$$$$$$$$$$$$$$$$$$$$$$$$$$$$$$$$$$$$$$$$$$$$$$$$$$$$$$$$$$$$$$$$$$$$$$$$$$$$$$$$$$$$$$$$$$$
%##################################################################################################################
%&&&&&&&&&&&&&&&&&&&&&&&&&&&&&&&&&&&&&&&&&&&&&&&&&&&&&&&&&&&&&&&&&&&&&&&&&

\section{Reconstruction and observational constraints}\label{s4}

\begin{figure}[h]
\begin{center}
{\includegraphics[scale=.3]{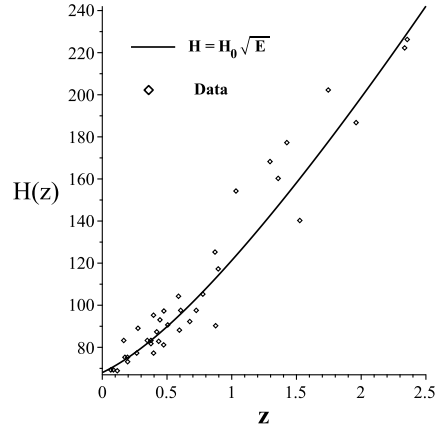}}
\caption{The Hubble parameter in terms of redshift parameter for $A_3 = 0.14$, $A_2 = 0.75$, $A_1 = -1.05$ and $A_0 = 1.16$.}\label{fig0}
\end{center}
\end{figure}

\begin{figure}[t]
\begin{center}
{\includegraphics[scale=.3]{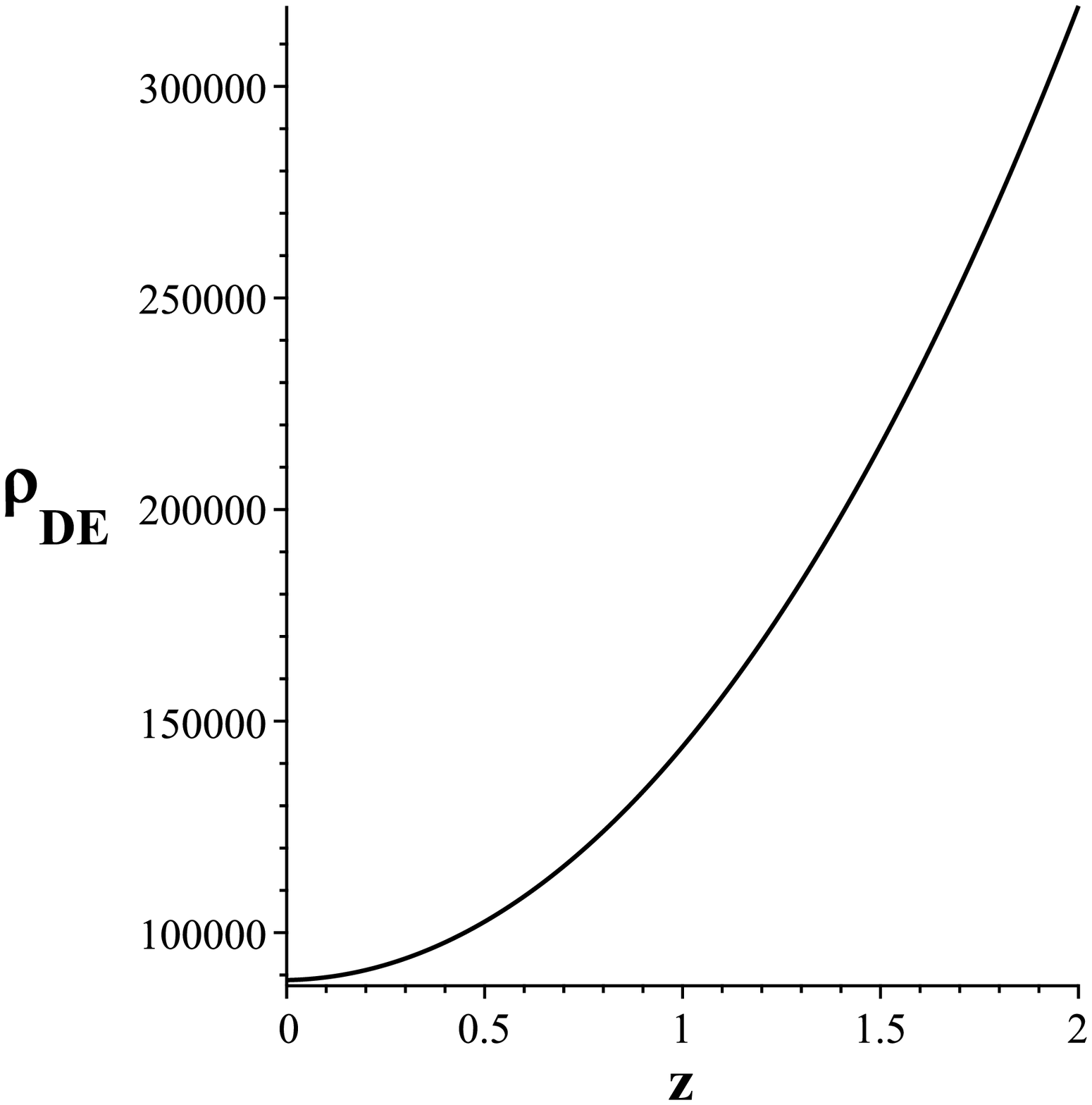}\label{fig1-1}}
{\includegraphics[scale=.3]{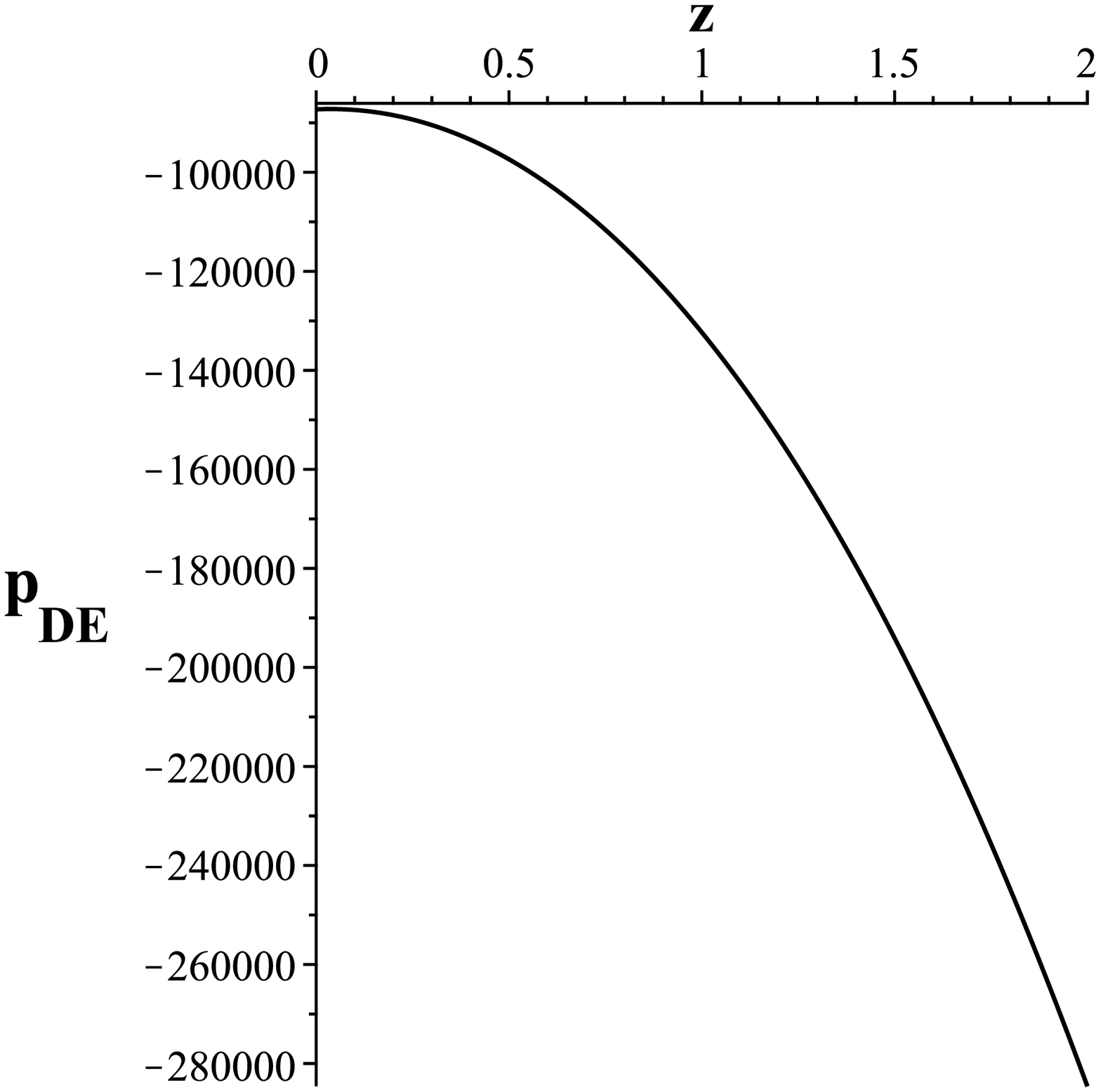}\label{fig1-2}}
\caption{The energy density and pressure of dark energy in terms of redshift parameter for $A_3 = 0.14$, $A_2 = 0.75$, $A_1 = -1.05$, $\alpha = 0.5$, $\beta = 0.47$, $\rho_{m_0} = 5$, $\omega_m = 0.05$, $a_0 = 1$ and $\xi = 0.75$.}\label{fig1}
\end{center}
\end{figure}

\begin{figure}[h]
\begin{center}
\subfigure
{\includegraphics[scale=.3]{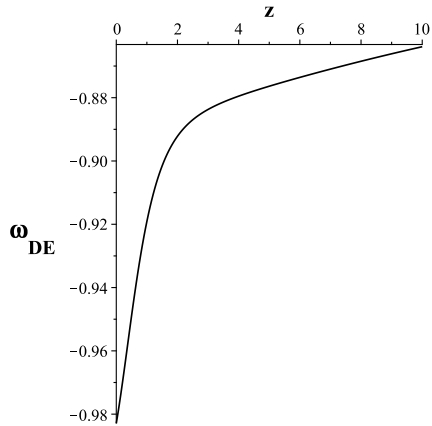}}
\caption{The EoS of dark energy in terms of redshift parameter for $A_3 = 0.14$, $A_2 = 0.75$, $A_1 = -1.05$, $\alpha = 0.5$, $\beta = 0.47$, $\rho_{m_0} = 5$, $\omega_m = 0.05$, $a_0 = 1$ and $\xi = 0.75$.}\label{fig2}
\end{center}
\end{figure}

\begin{figure}[h]
\begin{center}
\subfigure
{\includegraphics[scale=.3]{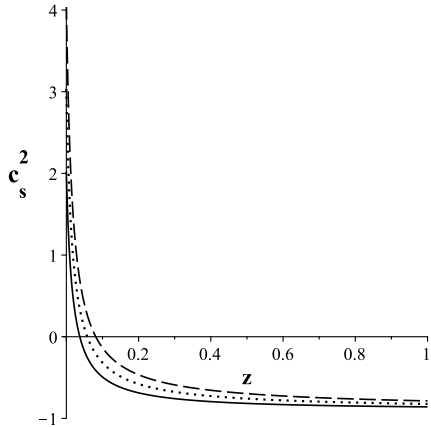}}
\caption{The graph $c_s^2$ in terms of redshift parameter for $A_3 = 0.14$, $A_2 = 0.75$, $A_1 = -1.05$, $\alpha = 0.5$, $\beta = 0.47$, $\rho_{m_0} = 5$, $\omega_m = 0.05$, $a_0 = 1$ and $\xi = 0.75\, \textrm{(line)}, 20\, \textrm{(dot)}, 40\, \textrm{(dash)}$.}\label{fig3}
\end{center}
\end{figure}

In this section, we intend to reconstruct the achievements of $f(T, \mathcal{T})$ gravity model by redshift parameter as well as survey by the observational constraints. For this purpose, we introduce the relationship between the scale factor and redshift parameter $z$ as $\frac{a_0}{a}=1+z$ in which $a_0$ demonstrates  the scale factor value in late time or $z = 0$. In that case, this equation leads to the differentially relationship as $\frac{d}{dt} = -H(1+z)\frac{d}{dz}$, which the time derivative relates to the redshift derivative. On the other hand, we parameterize the Hubble parameter by the parametrization function $E(z)$ as $H^2 = H_0^2 E(z)$ in which $H_0 = 68 \pm 2.8 \, km s^{-1} Mpc^{-1}$ is the Hubble parameter value at late time or $z = 0$. Nonetheless, we can rewrite the energy density and the pressure of dark energy \eqref{rhopdei} in terms of redshift parameter in the following form
\begin{subequations}\label{rhopdeii}
\begin{eqnarray}
 & \rho_{DE} = \frac{3 \alpha H_0^2}{\kappa^2 - 2 \beta} E - \frac{5 \alpha \beta H_0^2}{2 \left( {\kappa}^{2}+\beta \right)\left( {\kappa}^{2}-2 \beta \right)} (1+z) \partial_z E - \rho_{m_0} \left( \frac{1+z}{a_0}\right)^{3 (1+\omega_m)},\label{rhopdeii-1}\\
 &p_{DE} =  -\frac {3 \alpha H_0^2}{\kappa^{2}-2\,\beta} E + \frac {\alpha\, \left( 2 \kappa^{2}+\beta \right) H_0^2}{ 2 \left( \kappa^{2}+
\beta \right)  \left(\kappa^{2}-2\,\beta \right) } (1+z) \partial_z E + 3\,\xi\,H_0 \sqrt{E} - \omega_m \rho_{m_0} \left( \frac{1+z}{a_0}\right)^{3 (1+\omega_m)},\label{rhopdeii-2}
\end{eqnarray}
\end{subequations}
where the parametrization function $E(z)$ is determined by using the observational constraints. In order to evaluate the present system, we use the parametrization function in the following form \cite{Sahni_2003, Pourbagher_2019}
\begin{equation}\label{rz1}
E(z) =A_3 (1+z)^3 + A_2 (1+z)^2 + A_1 (1+z) + A_0,
 \end{equation}
 where $A_0$  to $A_3$ are constant and their values are obtained by fitting the observational data as $A_3 = -0.16 \pm 0.30$, $A_2 =  2.39 \pm 1.71$, $A_1= -3.80 \pm 2.89$ and $A_0 = 1-A_3-A_2-A_1$ in which with condition $E(z=0)=1$  (i.e. parametrization function is equal one in current universe ($z = 0$)) we find $A_3+A_2+A_1+A_0=1$. It is interesting to note that this function is inspired by Ref. \cite{Sahni_2003} which corresponds to the observational data. In that case, we plot the variety of the Hubble parameter versus the redshift parameter by the aforesaid parametrization function as shown in Fig. \ref{fig0}. Fig. \ref{fig0} shows us that the parametrization function is compatible  with the observational data by an excellent fitting.
 We note that in this study the choice of free parameters is very sensitive and play a key role for describing the accelerated extension of the universe. Thus, their choice motivation is that the graphs of the energy density and the pressure versus the redshift parameter are positive and negative, respectively. By this approach, we draw graphs of the energy density and the pressure of dark energy in terms of redshift parameter by values $A_3 = 0.14$, $A_2 = 0.75$, $A_1 = -1.05$, $\alpha = 0.5$, $\beta = 0.47$, $\rho_{m_0} = 5$, $\omega_m = 0.05$, $a_0 = 1$ and $\xi = 0.75$ according to Fig. \ref{fig1}.\\
As we know, we can specify the accelerated expansion of the universe with a parameter called the EoS of dark energy. Thus, to obtain the EoS of dark energy, we insert the Eqs. \eqref{rhopdeii} into the Eq. \eqref{eos1}, and then the variation of EoS is drawn in terms of the redshift parameter as Fig. \ref{fig2}. But observations show that the present value of the EoS is close to $-1$ \cite{Amanullah_2010, Suzuki_2012, Aghanim_2018}. Therefore, Fig. \ref{fig2} shows us that the value of EoS is equal $-0.983$ for late time ($z = 0$), and one tells us that the universe is in a phase of accelerated expansion.\\
As we know, examining the stability and instability of a system is one of the most important foundations of that system. Therefore, Our homogeneous and isotropic universe needs to be studied for its stability. The physical meaning is that instability of a gravitational fluid versus small perturbations lead to growing non-controlled of energy density. So for a more complete description, we examine our findings from a thermodynamic viewpoint for a stability condition. For this purpose, we consider that the universe is in an adiabatic system, namely the entropy perturbation is zero ($\delta S=0$) and the pressure perturbs only with the energy density as $\delta p_{DE} = \frac{\partial p_{DE}}{\partial \rho_{DE}} \delta \rho_{DE}$ in which $c_s^2=\frac{\partial p_{DE}}{\partial \rho_{DE}}=\frac{\partial_z p_{DE}}{\partial_z \rho_{DE}} $ is defined as sound speed parameter. Thus the sound speed parameter is a determinative quantity for the system stability which  the condition $c_s^2 > 0$ represents stability and the condition $c_s^2 < 0$ indicates instability. With this interpretation, we plot the graph $c_s^2 $ versus the redshift parameter by different values of viscosity coefficient as shown in Fig. \ref{fig3}. The Fig. \ref{fig3} displays that the value of the sound speed parameter in late time ($z = 0$) is equal to $2.023$, $3.035$ and $4.088$ for $\xi = 0.75, 20, 40$, respectively. This means the corresponding system is undergoing a stability condition in late time and is instability in early time.

%$$$$$$$$$$$$$$$$$$$$$$$$$$$$$$$$$$$$$$$$$$$$$$$$$$$$$$$$$$$$$$$$$$$$$$$$$$$$$$$$$$$$$$$$$$$$$$$$$$$$$$$$$$$$$$$$$$$$$$$$$$$
%##################################################################################################################
%&&&&&&&&&&&&&&&&&&&&&&&&&&&&&&&&&&&&&&&&&&&&&&&&&&&&&&&&&&&&&&&&&&&&&&&&&

\section{Conclusion}\label{s5}
In this paper, the $f(T,\mathcal{T})$ gravity model has been studied as one of teleparallel gravity scenarios in which $T$ and $\mathcal{T}$ are torsion scalar and trace of the matter energy-momentum tensor, respectively. For this purpose, to use the Weitzenb\"{o}ck connection rather than the Levi-Civita connection, the field equation has obtained by the tetrad field in the flat-FRW background. We considered that there is a viscous fluid called bulk viscosity in within the universe and then obtained the Friedmann equations in terms of the effective energy density and the effective pressure. The universe components have been considered a combination of dark energy and matter, and then to use the Friedmann equations, the energy density and the pressure of dark energy have obtained. In what follows, to use the parametrization approach $E = \frac{H^2}{H_0^2}$, the energy density and the pressure of dark energy have written in terms of the redshift parameter. Then, we took the parametrization function as $E(z) =A_3 (1+z)^3 + A_2 (1+z)^2 + A_1 (1+z) + A_0$ which one has best fitted with astronomical data. In order to solve the present system, we have considered the arbitrary function $f(T,\mathcal{T})$ as a linear combination of $T$ and $\mathcal{T}$. Then we plotted the cosmological parameters in terms of the redshift parameter, and the graphs showed us that the universe is undergoing a phase of accelerated expansion. Finally, we explored the system stability by a function called the speed sound function, which the achieved result showed us that there is stability in late time.

%$$$$$$$$$$$$$$$$$$$$$$$$$$$$$$$$$$$$$$$$$$$$$$$$$$$$$$$$$$$$$$$$$$$$$$$$$$$$$$$$$$$$$$$$$$$$$$$$$$$$$$$$$$$$$$$$$$$$$$$$$$$
%##################################################################################################################
%&&&&&&&&&&&&&&&&&&&&&&&&&&&&&&&&&&&&&&&&&&&&&&&&&&&&&&&&&&&&&&&&&&&&&&&&&

\end{document}